\documentclass[runningheads]{llncs}
\usepackage[T1]{fontenc}
\usepackage{graphicx}
\usepackage{subcaption} 
\usepackage{booktabs}
\usepackage{multirow}
\usepackage{orcidlink}

\begin{document}
\title{Silicon Showdown: Performance, Efficiency, and Ecosystem Barriers in Consumer-Grade LLM Inference}
\titlerunning{\textit{Silicon Showdown}}
%
\author{Abdurrahman Javat\orcidlink{0009-0004-5024-0376} ~\and 
Allan Kazakov\orcidlink{0009-0002-2122-0064}}
\authorrunning{\textit{A. Javat and A. Kazakov}}
%
\institute{Bahcesehir University, Istanbul, T\"urkiye 
\\ \email{\{abdurrahman.javat,allan.kazakov\}@bahcesehir.edu.tr}}

%
\maketitle              
\begin{abstract}
The operational landscape of local Large Language Model (LLM) inference has shifted from lightweight models to datacenter-class weights exceeding 70B parameters, creating profound systems challenges for consumer hardware. This paper presents a systematic empirical analysis of the Nvidia and Apple Silicon ecosystems, specifically characterizing the distinct intra-architecture trade-offs required to deploy these massive models. On the Nvidia Blackwell architecture, we identify a critical "Backend Dichotomy" within the TensorRT-LLM stack: while the new NVFP4 quantization format delivers a 1.6x throughput advantage over optimized BF16 baselines (151 tokens/s vs. 92 tokens/s), realizing this performance requires navigating complex runtime constraints that trade startup latency for generation speed. Furthermore, we characterize the "VRAM Wall" for 70B+ models: on discrete GPUs, users face a destructive choice between aggressive quantization (e.g., Q2) that degrades model intelligence to fit in VRAM, or PCIe-bottlenecked CPU offloading, which reduces throughput by over 90\% compared to full-GPU execution. Conversely, Apple’s Unified Memory Architecture (UMA) circumvents these bottlenecks, enabling linear scaling for 80B parameter models at practical 4-bit precisions. This architectural divergence extends to operational sustainability, where Apple’s SoC design demonstrates up to a 23x advantage in energy efficiency (tokens/joule). We conclude that for consumer-grade inference, the optimal hardware is defined by a complex interplay between compute density (Nvidia) and memory capacity (Apple), moderated by the significant "ecosystem friction" of proprietary quantization workflows.

\end{abstract}

\section{Introduction}
\label{sec:introduction}
The operational landscape of local Large Language Model (LLM) inference has fundamentally shifted. While early adoption was characterized by lightweight, dense models, the proliferation of high-capacity weights and the rapid adoption of Mixture-of-Experts (MoE) architectures \cite{switch-transformer} has created a new imperative: executing datacenter-class reasoning on consumer workstations. This transition, fueled by open-weight releases from labs like DeepSeek \cite{deepseek} and Meta \cite{meta-llama}, has transformed local inference from a novelty into a viable alternative to closed-source APIs. However, deploying 70B+ parameter models locally introduces profound systems challenges. It places the consumer hardware ecosystem in a vice, squeezing it between the Memory Capacity required to retain these massive parameter sets and the Compute Density required to generate tokens at interactive latencies.

The current hardware landscape offers two divergent solutions to this problem, each defined by a distinct architectural philosophy. Nvidia’s discrete GPUs (e.g., the Blackwell-based RTX 5090) prioritize massive parallel throughput and specialized tensor acceleration. With the introduction of the NVFP4 quantization format \cite{nvfp4}, these devices promise near-FP16 accuracy at 4-bit speeds, establishing a new theoretical ceiling for consumer inference. However, they remain constrained by the "VRAM Wall"—a hard physical limit on video memory (e.g., 32GB) that forces complex offloading strategies for larger models. In contrast, Apple’s M-series SoCs employ a Unified Memory Architecture (UMA), granting the GPU direct access to the entire system memory pool (up to 512GB). This design eliminates the PCIe bottleneck, offering a linear scaling path for model capacity that discrete architectures struggle to match.

While the hardware dictates the potential, the software ecosystem determines the reality. We posit that raw throughput metrics (tokens/second) are no longer sufficient to evaluate consumer inference. A true systems-level analysis must quantify the "ecosystem friction" involved in deployment. On the Nvidia platform, the transition to TensorRT-LLM v1.1.0 and NVFP4 has introduced a labyrinth of trade-offs: the choice between PyTorch and \textit{C++} backends can halve latency at the cost of throughput, while the compilation of engines for large models often exceeds the memory capacity of the GPU itself. Conversely, Apple’s MLX framework offers a "native" path, leveraging dynamic graph construction to run 4-bit quantized models without ahead-of-time compilation.

To rigorously evaluate these trade-offs, we present a systematic empirical study of high-end consumer inference, ranging from efficient 1B models to massive 80B parameter workloads. We replace previous evaluations of legacy Int4/Int8 formats with an analysis of the modern quantization frontier: NVFP4 on Blackwell, GGUF CPU-offloading, and MLX 4-bit native quantization.

Our key contributions are as follows:
\begin{enumerate}
\item \textbf{We quantify the "NVFP4 Advantage" and the "Backend Dichotomy":} On the Nvidia RTX 5090, the NVFP4 format delivers a 1.6x throughput increase over standard BF16 baselines. However, we uncover a critical instability within TensorRT-LLM v1.1.0: users must choose between a PyTorch backend (151 tok/s, 31ms startup) and a \textit{C++} backend (93 tok/s, 19ms startup), complicating deployment.
\item \textbf{We identify "Ecosystem Friction" as a primary deployment barrier:} We demonstrate that while both MLX and TensorRT-LLM support local quantization, the resource requirements diverge drastically. We find that quantizing even a modest 8B parameter model into NVFP4 exceeds the 32GB VRAM capacity of the flagship RTX 5090, forcing users to rely on pre-quantized weights published by Nvidia on HuggingFace (the list of the models is limited) or slow CPU offloading with GGUF. Furthermore, we observe a gap between the release of new open-source models and their availability in optimized inference stacks. For example, recent architectures such as GLM-4.7-Flash have not yet been integrated into TensorRT-LLM, which may require practitioners to rely on alternative frameworks like llama.cpp \cite{llama.cpp}, where certain NVIDIA-specific optimizations (e.g., NVFP4 quantization) are not currently supported.
\item \textbf{We characterize the "70B+ VRAM Wall":} We demonstrate that for moderately sized, state-of-the-art open-source models (70B-80B parameters), the memory limit of discrete GPUs forces a destructive choice: utilize aggressive quantization (e.g., Q2/IQ3) that degrades model intelligence, or offload layers to the CPU via PCIe, which collapses inference speed.
\item \textbf{We quantify the "Performance-Efficiency Continuum" and architectural maturity gaps:} Using a lightweight 1.5B baseline, we reveal a stark architectural dichotomy: while the RTX 5090 delivers 70\% higher throughput, the Apple M3 Ultra achieves a staggering 23x advantage in energy efficiency. Furthermore, we identify a "Software Maturity Gap" in the Blackwell architecture, where the previous-generation RTX 4090 outperforms the flagship RTX 5090 by more than 2.2x in Time-To-First-Token (TTFT) latency, highlighting a possible software optimization lag for the newest hardware generation.
\end{enumerate}

\section{Background and Related Work}
\label{sec:background}

To contextualize our findings, we first examine the foundational architectural differences between the hardware platforms and the operational principles of their respective acceleration frameworks and quantization formats.

\subsection{Architectural Foundations: Discrete vs. Unified Memory}
The consumer hardware landscape for high-performance computing is primarily defined by two competing memory architectures: the discrete model, epitomized by Nvidia's RTX-series GPUs, and the unified model, championed by Apple's M-series SoCs.

\textbf{Discrete GPU Architecture}: Nvidia's GPUs act as specialized co-processors connected via a PCIe bus, featuring a dedicated pool of high-bandwidth GDDR VRAM. This memory enables the massive parallel throughput required for NVFP4 \cite{nvfp4} operations. However, this architecture imposes a "VRAM Wall": the capacity is fixed at the hardware level (e.g., 32GB on the RTX 5090). For models exceeding this limit, such as Llama-3.3-70B, data must be offloaded to system RAM and transferred over the PCIe bus during inference, or quantized heavily to fit completely on the GPU. Offloading introduces a significant bottleneck, as PCIe bandwidth (typically 32-64 GB/s) is an order of magnitude slower than GDDR bandwidth (approx. 1,800 GB/s).

\textbf{Unified Memory Architecture (UMA)}: Apple Silicon SoCs employ a design where the CPU and GPU share a single, massive pool of LPDDR memory. In this UMA model, there is no separate VRAM and no PCIe bottleneck. A key advantage for LLMs is flexibility; the GPU can access the entire system memory (up to 512GB), allowing it to run light quantized 70B+ parameter models without the latency penalties associated with offloading on discrete architectures.

\subsection{Software Acceleration and Quantization Ecosystems}
Hardware capabilities are unlocked by software. We analyze three distinct approaches to deploying LLMs on consumer hardware:

\textbf{TensorRT-LLM and NVFP4 (Nvidia)}: TensorRT-LLM is an Ahead-of-Time (AOT) compilation framework that builds optimized "engines" for specific GPUs. Version 1.1.0 introduces two critical changes. First, it supports \textbf{NVFP4}, a 4-bit floating-point format specific to the Blackwell architecture. Unlike integer quantization (Int4), NVFP4 retains high dynamic range, offering near-FP16 accuracy with significant memory savings\cite{nvfp4}. Second, the framework now bifurcates its runtime into two backends: a C++ Runner (legacy, optimized for latency) and a PyTorch Runner (default, optimized for throughput). Users could explicitly choose between these backends, often trading startup speed for token generation rate.

\textbf{MLX (Apple Silicon)}: MLX is a framework designed for Apple’s UMA. It employs lazy evaluation and dynamic graph construction, avoiding the rigid AOT compilation step of TensorRT-LLM. MLX supports native 4-bit quantization that operates directly on the unified memory pool, allowing users to load and quantize models dynamically without the high-VRAM requirements of compiling a static engine.

\textbf{GGUF and CPU Offloading}: The GGUF format, popularized by the `llama.cpp` project \cite{llama.cpp}, is the standard fallback for hardware-constrained inference. It utilizes a heterogeneous computing model: layers that fit in VRAM are computed on the GPU, while the remainder are offloaded to the CPU and system RAM. This allows users to run models larger than their GPU’s VRAM, albeit at speeds limited by the PCIe bus transfer rate.

\subsection{Related Work in LLM Benchmarking}
Our work builds upon industry standards like MLPerf \cite{Reddi2020MLPerf}, which established reproducible methodology for datacenter inference. However, MLPerf and academic studies on systems like vLLM \cite{Kwon2023vLLM} primarily focus on maximizing batch throughput in multi-tenant server environments with unlimited power budgets (e.g., Nvidia H100s).

Conversely, the domain of high-end consumer inference "local workstations" remains under-explored. Existing analyses \cite{PugetSystems2024} often focus solely on throughput (tokens/sec) for models that fit entirely in VRAM. They fail to quantify the "cliff" that occurs when models exceed VRAM capacity, nor do they analyze the ecosystem friction of modern quantization formats. To the best of our knowledge, this is the first systematic study comparing the \textbf{NVFP4} quantization frontier against \textbf{Unified Memory} scaling for datacenter-class models (70B+) on consumer hardware.

\section{Methodology}
\label{sec:methodology}

To conduct a fair and reproducible comparison of the Nvidia and Apple Silicon ecosystems, we designed a systematic benchmarking protocol. This protocol was executed across a suite of high-end consumer hardware to capture the performance characteristics of each architecture at the frontier of local inference. This section details the hardware and software configurations, the specific quantization strategies employed, and the metrics collected.

\subsection{Hardware and Software Configurations}
\label{ssec:hw-sw-config}

Our study evaluates hardware configurations representing the current flagship capabilities of both ecosystems. The specifications for these systems are detailed in Table \ref{tab:nvidia-hardware} and Table \ref{tab:apple-hardware}.

\begin{table}[h]
\caption{Nvidia hardware configurations} 
\label{tab:nvidia-hardware}
\centering
\footnotesize 
\setlength{\tabcolsep}{8pt} 
\begin{tabular}{lcccccc}
\toprule
\textbf{Spec} & \textbf{5090} & \textbf{5080} & \textbf{4090} & \textbf{4050} & \textbf{4050} & \textbf{3050} \\
& & & & & \textit{(Lap)} & \textit{(Lap)} \\ 
\midrule
Transistors (B) & 92.2 & 45.6 & 76.3 & 18.9 & 18.9 & 8.7 \\
VRAM (GB) & 32 & 16 & 24 & 8 & 6 & 4 \\
Bandwidth (GB/s) & 1792 & 960 & 1008 & 272 & 192 & 192 \\
CUDA Cores & 21760 & 10752 & 16384 & 3072 & 2560 & 2048 \\
Tensor Cores & 680 & 336 & 512 & 96 & 80 & 64 \\
L2 Cache (MB) & 96 & 64 & 72 & 24 & 24 & 2 \\ 
\bottomrule
\multicolumn{7}{l}{\emph{Note: All Nvidia testbeds run Ubuntu 24.04 LTS.}}
\end{tabular}
\end{table}

\begin{table}[h]
\caption{Apple Silicon hardware configurations} 
\label{tab:apple-hardware}
\centering
\footnotesize
\setlength{\tabcolsep}{8pt}
\begin{tabular}{lcccccc}
\toprule
\textbf{Spec} & \textbf{M3} & \textbf{M4} & \textbf{M2} & \textbf{M2} & \textbf{M2} & \textbf{M1} \\
& \textbf{Ultra} & \textbf{Pro} & \textbf{Max} & \textit{(Air)} & \textit{(Pro)} & \\ 
\midrule
Transistors (B) & 184.0 & 28.0 & 67.0 & 20.0 & 20.0 & 16.0 \\
Memory (GB) & 96 & 64 & 32 & 16 & 8 & 8 \\
Bandwidth (GB/s) & 819.2 & 273.0 & 409.6 & 102.4 & 102.4 & 68.3 \\
CPU Cores & 28 & 14 & 12 & 8 & 8 & 8 \\
GPU Cores & 60 & 20 & 38 & 8 & 8 & 8 \\
Cooling\textsuperscript{a} & Act. & Act. & Act. & Pas. & Act. & Pas. \\ 
\bottomrule
\multicolumn{7}{l}{\textsuperscript{a} Act. = Active (fan); Pas. = Passive (fanless/silent).}
\end{tabular}
\end{table}

All benchmarks were conducted using a standardized software stack. The Nvidia testbeds utilized Ubuntu 24.04. This Linux environment was a prerequisite for the TensorRT-LLM framework. The Apple Silicon devices ran macOS 26 Tahoe.

For the accelerated benchmarks, we utilized \textbf{TensorRT-LLM v1.1.0} \cite{trt-llm} on the Nvidia platform and \textbf{MLX v0.30.6} \cite{mlx} on Apple Silicon. For hybrid CPU-offloading scenarios on Nvidia, we utilized \texttt{llama.cpp} (GGUF) \cite{llama.cpp} using the Unsloth's model repositories on HuggingFace. Notably, for the TensorRT-LLM benchmarks (model size $<10$B), we explicitly utilized the framework's \textbf{PyTorch backend} rather than the C++ runner, as preliminary testing indicated superior throughput stability.

\subsection{Benchmark Models and Quantization Strategies}
\label{ssec:models-workloads}

Our study focuses on two distinct model categories: high-efficiency small models to test compute density, and datacenter-class large models (70B-80B) to test memory capacity. The selected models and their specific precision configurations for the Nvidia benchmarks are detailed in Table \ref{tab:benchmark-models}.

\begin{table}[h]
\centering
\caption{Benchmark language models} 
\label{tab:benchmark-models}
\footnotesize 
\setlength{\tabcolsep}{4pt}
\begin{tabular}{lcl}
\toprule
\textbf{Model Name} & \textbf{Params} & \textbf{Precision} \\ 
\midrule
Qwen2.5-1.5B \cite{qwen-2.5} & 1.5B & float16 \\
Qwen3-8B \cite{qwen} & 8B & bfloat16, NVFP4(PyTorch \& C++) \\
GLM4.7-Flash \cite{glm-4.5} & 30B-A3B MoE & Q6\_K\_XL(GGUF) \\
Llama-3.3-70B-Instruct \cite{meta-llama} & 70B & Q4\_K\_M, IQ3\_XXS (both GGUF) \\
Qwen3-Next-80B-A3B \cite{qwen} & 80B-A3B MoE & Q4\_K\_M, Q2\_K\_XL (both GGUF) \\
\bottomrule
\end{tabular}
\end{table}

\subsubsection{Quantization on Nvidia (Blackwell):}
For the Qwen3-8B model on the RTX 5090, we utilized the \textbf{NVFP4} quantization format. As documented by Nvidia \cite{nvfp4}, NVFP4 is now the de-facto standard for inference on the Blackwell architecture, delivering throughput comparable to legacy Int4 while maintaining accuracy slightly below native BFloat16. We note that while advanced variants such as W4A8 (NVFP4 weights with FP8 activations) are currently under development and testing, they are not supported in the stable TensorRT-LLM v1.1.0 release used for this study. Therefore, all NVFP4 benchmarks utilized NVFP4 weights with standard FP16 activations.

\subsubsection{The GGUF Offloading Dilemma:}
For models exceeding the 32GB VRAM capacity of consumer flagships (e.g., Llama-3.3-70B), users face a critical trade-off. To capture this, we benchmarked two distinct GGUF configurations reflecting real-world usage:
\begin{enumerate}
    \item \textbf{Maximal Accuracy (Offloading):} We utilized standard quantization (e.g., \texttt{Q4\_K\_M}). Since these models exceed VRAM, the amount of offloaded layers was chosen to maximize VRAM utilization, and the rest were offloaded to the CPU/System RAM, prioritizing model intelligence over speed.
    \item \textbf{Maximal Throughput (Full GPU):} We utilized aggressive quantization (e.g., \texttt{IQ3\_XXS} or \texttt{Q2\_K\_XL}) specifically chosen to fit the model entirely within the VRAM, eliminating the PCIe bottleneck but potentially degrading generation quality.
\end{enumerate}

\subsubsection{Quantization on Apple Silicon:}
For all models listed in Table \ref{tab:benchmark-models} with parameters $\ge$ 30B, the Apple Silicon benchmarks utilized \textbf{native 4-bit MLX quantization}. Unlike the discrete GPU ecosystem, Apple's Unified Memory Architecture allows even the 80B parameter model to fit comfortably within the 96GB memory of the M3 Ultra, leaving no reason to test the llama.cpp GGUF formats on Apple Silicon as it has been empirically shown that MLX provides the best throughput for LLM inference on Apple Silicon devices \cite{mlx-comparison}.

\subsection{Experimental Design and Metrics}
\label{ssec:metrics}

To ensure the isolation of architectural performance characteristics from transient system noise, we implemented a rigorous experimental protocol.

\textbf{Workload Configuration:}
The inference workload was constructed to stress both the compute-bound "prefill" phase and the memory-bound "decode" phase. We utilized a standardized dataset of prompts categorized into three distinct input lengths: Short ($\approx$16 tokens), Medium ($\approx$300 tokens), and Long ($\approx$496 tokens). To normalize the comparison of generation speed, all benchmarks were configured to produce a fixed output length of 256 tokens using greedy decoding. 

\textbf{Experimental Control:}
To mitigate the impact of dynamic frequency scaling (DFS) and thermal throttling, each benchmark run was preceded by a "warm-up" phase consisting of two untimed generation cycles. This ensures the GPU clock speeds have stabilized prior to data collection. Furthermore, a fixed random seed was enforced across all frameworks (PyTorch, TensorRT-LLM, MLX) to guarantee deterministic token generation paths.

\textbf{Metrics Collected:}
We report the average of three consecutive timed runs for the following metrics:
\begin{itemize}
    \item \textbf{Generation Throughput (Tokens/s):} Calculated strictly during the auto-regressive decode phase. This metric isolates the memory bandwidth performance of the system.
    \item \textbf{Time-To-First-Token (TTFT):} The wall-clock latency (ms) from request submission to the appearance of the first token. This metric isolates the massive parallel compute capability of the GPU during the prompt processing (prefill) phase.
    \item \textbf{Energy Efficiency (Tokens/Joule):} 
    On Nvidia platforms, we polled the \texttt{PyNVML} energy counters at 100ms intervals. 
    On Apple Silicon, we utilized the \texttt{powermetrics} utility to sample SoC power draw. 
    Total energy was derived by integrating power consumption over the duration of the inference task ($E = \int_{t_{start}}^{t_{end}} P(t) dt$).
\end{itemize}

\section{Results}
\label{sec:results}

Our empirical evaluation reveals a multi-faceted performance landscape defined by architectural trade-offs, software ecosystem maturity, and the physical constraints of memory capacity. We begin by analyzing the fundamental performance-efficiency continuum established by small models, before analyzing the specific advantages of Nvidia's new NVFP4 precision format and the capacity challenges posed by state-of-the-art 70B+ parameter models.

\subsection{The Performance-Efficiency Continuum}
\label{ssec:frontier}

We begin our analysis by examining the core architectural characteristics of the two platforms using the Qwen2.5-1.5B model. This lightweight workload serves as a baseline to isolate the raw throughput and latency characteristics of the hardware without the memory capacity bottlenecks that characterize larger models.

\textbf{The Architectural Dichotomy:} Our data reveals two distinct operational clusters. Nvidia devices establish a high-power, high-performance frontier, while Apple Silicon devices define a low-power, high-efficiency alternative. Using the optimized TensorRT-LLM framework, the flagship RTX 5090 delivers a peak throughput of 265 tokens/second, a 70\% increase over the 155 tokens/second achieved by the Apple M3 Ultra with MLX. 

However, this raw performance advantage comes at a substantial energy cost. While we exclude specific energy plots for brevity, our measurements indicate a dramatic inversion in efficiency: the Apple M3 Ultra delivers up to 23 times more tokens per joule than the RTX 5090. This stems directly from the System-on-Chip (SoC) design, where the Unified Memory Architecture (UMA) minimizes power-hungry data transfers between discrete components, whereas the discrete GPU architecture prioritizes maximum clock speeds and memory bandwidth at the expense of power draw.

\begin{figure*}[t!]
\centering
\begin{subfigure}{0.48\columnwidth}
    \centering
    \includegraphics[width=\linewidth]{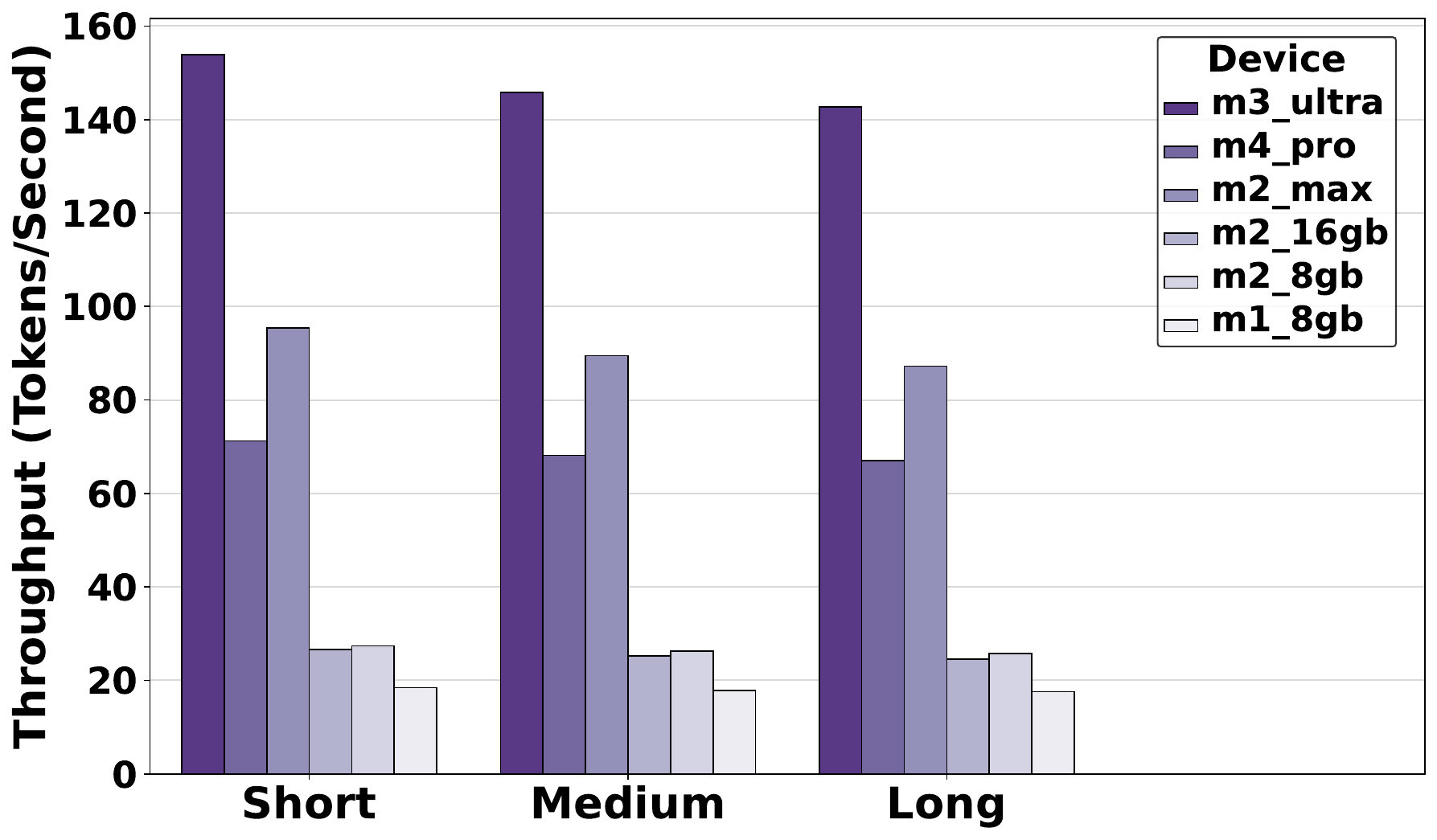}
    \caption{Apple Silicon Throughput (Tok/s)}
    \label{fig:apple-throughput}
\end{subfigure}
\hfill
\begin{subfigure}{0.48\columnwidth}
    \centering
    \includegraphics[width=\linewidth]{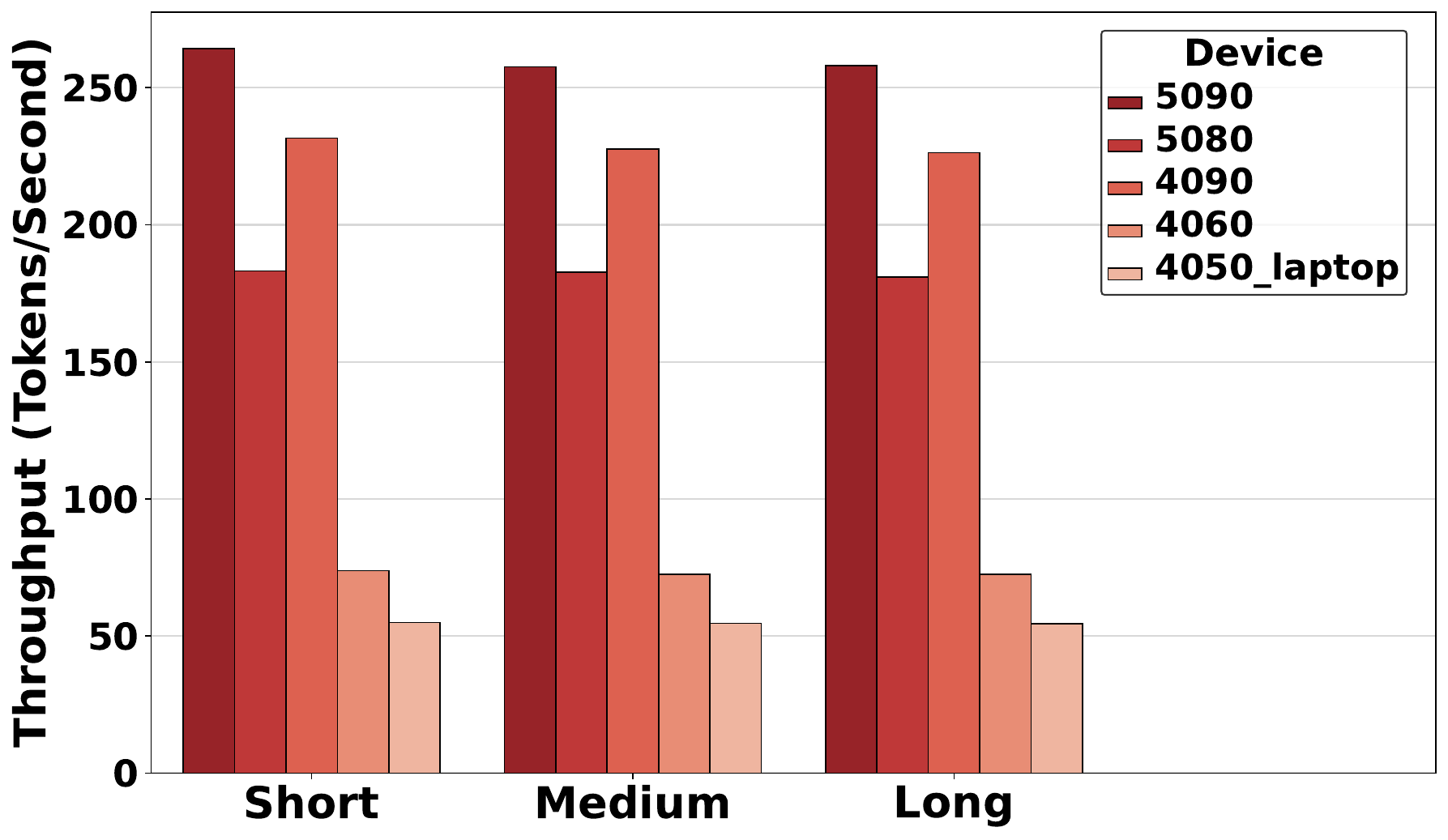}
    \caption{Nvidia Throughput (Tok/s)}
    \label{fig:nvidia-throughput}
\end{subfigure}

\vspace{1em} 

\begin{subfigure}{0.48\columnwidth}
    \centering
    \includegraphics[width=\linewidth]{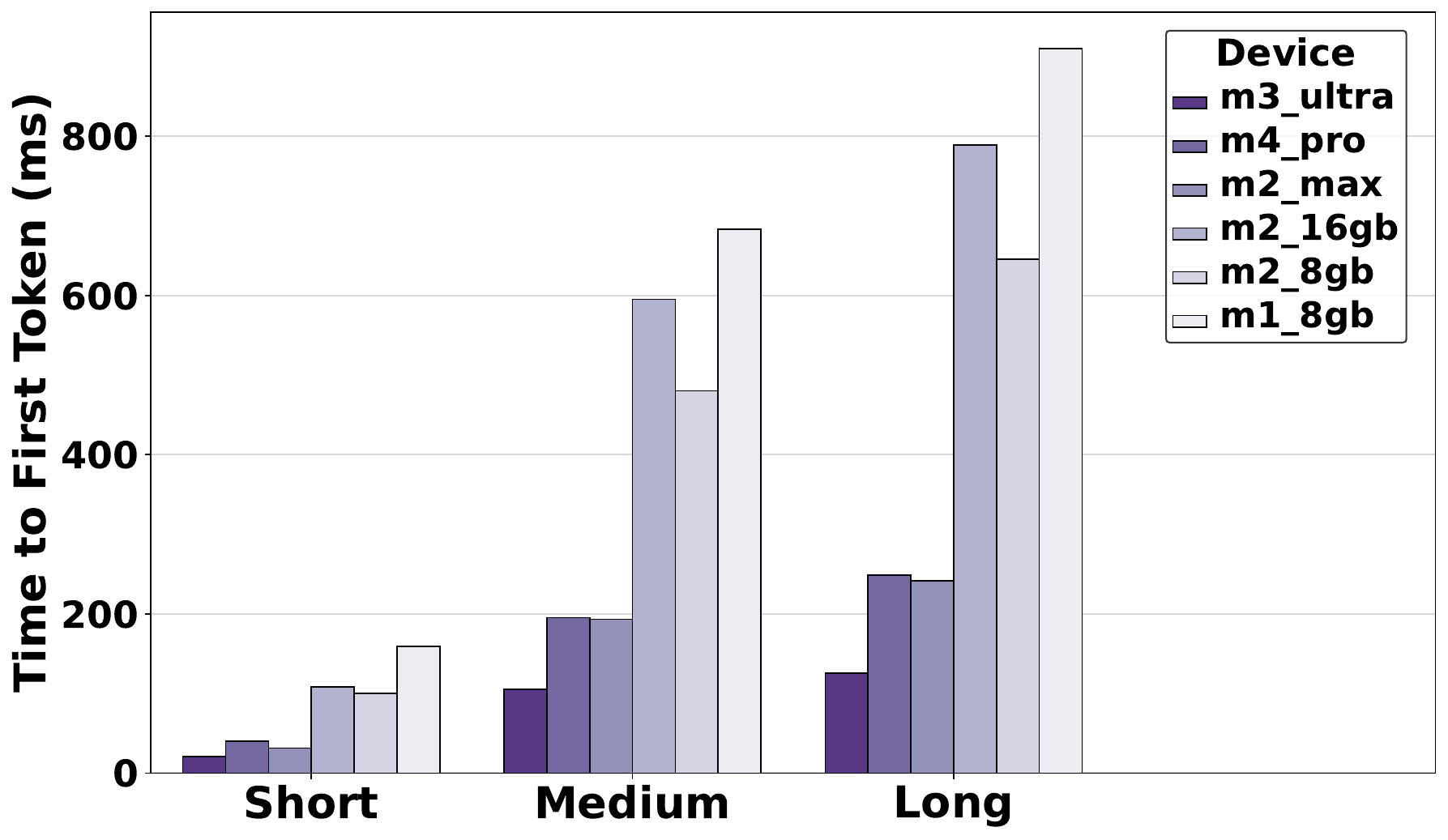}
    \caption{Apple Silicon TTFT (ms)}
    \label{fig:apple-ttft}
\end{subfigure}
\hfill
\begin{subfigure}{0.48\columnwidth}
    \centering
    \includegraphics[width=\linewidth]{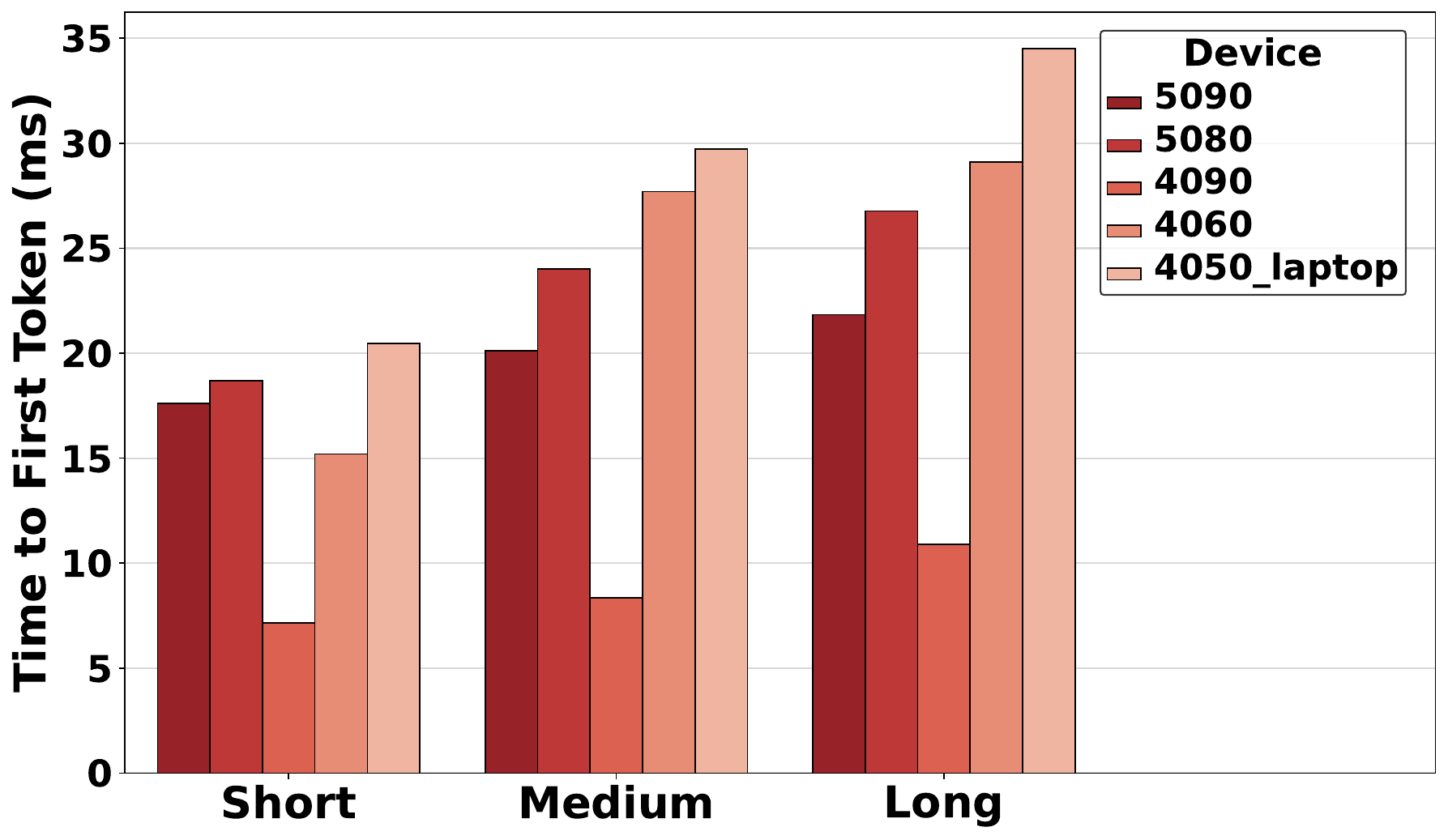}
    \caption{Nvidia TTFT (ms)}
    \label{fig:nvidia-ttft}
\end{subfigure}

\caption{Intra-Architecture Performance Grid for Qwen2.5-1.5B. (a) Apple throughput scales with core count, showing distinct tiers between Air (M2) and Pro/Ultra chips. (b) Nvidia throughput scales linearly with generation. (c) Apple latency is sensitive to thermal throttling (M2 Air vs Pro). (d) Nvidia reveals a software maturity gap, where the older RTX 4090 significantly outperforms the flagship RTX 5090 in startup latency.}
\label{fig:intra-arch-grid}
\end{figure*}

\textbf{Intra-Architecture Dynamics:} To understand how performance scales within each ecosystem, Figure~\ref{fig:intra-arch-grid} presents a 2x2 comparison of Throughput and Time-To-First-Token (TTFT) across the hardware generations.

On the Apple Silicon side (Figure~\ref{fig:intra-arch-grid}a and \ref{fig:intra-arch-grid}c), we observe a nuanced interplay between thermal management and sustained performance. Comparing the passively cooled 16GB M2 Air against the actively cooled 8GB M2 Pro reveals that while burst performance is similar, the active cooling solution allows the M2 Pro to sustain higher clock speeds, resulting in lower latency and higher aggregate throughput on longer contexts.

On the Nvidia side (Figure~\ref{fig:intra-arch-grid}b and \ref{fig:intra-arch-grid}d), raw throughput scales predictably with hardware generations, with the 50-series leading the pack. However, we identify a significant anomaly in latency. As shown in Figure~\ref{fig:intra-arch-grid}d, the previous-generation RTX 4090 consistently delivers the lowest TTFT (8ms), outperforming the flagship RTX 5090 (18ms) by more than 2.2x. 

This counter-intuitive regression is likely attributable to software maturity rather than hardware limitation. The RTX 5090 (Blackwell) represents a new architectural paradigm, and the current TensorRT-LLM software stack is in an active phase of optimization for this platform. While the drivers are not yet fully matured for the massive parallelism of the 50-series pre-fill stage, the hardware possesses architectural advantages, specifically regarding new precision formats, that we explore in the following subsection.

\subsection{The Compute Frontier: NVFP4 and the Backend Dichotomy}
\label{ssec:nvfp4}

With the introduction of the Blackwell architecture, Nvidia positioned \textbf{NVFP4} as the new de-facto standard for inference \cite{nvfp4}, promising the accuracy of BF16 with the computational density of INT4. We benchmarked the \textbf{Qwen3-8B} model on an RTX 5090 to evaluate this claim.

Our results, summarized in Table~\ref{tab:nvfp4-results}, reveal a fractured ecosystem. While NVFP4 is capable of massive performance uplifts, these gains are strictly gated behind the new "PyTorch-first" architecture. The legacy C++ engine, previously the gold standard for deployment, fails to utilize the format effectively, creating a stark "Backend Dichotomy."

\textbf{1. The Throughput Victory (PyTorch Backend):}
When utilizing the modern PyTorch backend (TensorRT-LLM v1.1.0 default), NVFP4 behaves as advertised. It achieves a throughput of \textbf{151.4 tokens/sec}, representing a \textbf{64\% performance increase} over the optimized GGUF BF16 baseline (92.2 tokens/sec). This confirms that with the correct runtime, NVFP4 successfully unlocks the bandwidth advantages of 4-bit weights on consumer hardware.

\textbf{2. The Legacy Optimization Failure:}
However, the Legacy C++ backend reveals a critical optimization gap. Despite running the same NVFP4 model, the Legacy engine collapses to \textbf{93.1 tokens/sec}, which is statistically identical to the optimized BF16 GGUF baseline.
While the Legacy backend offers a marginal improvement in Time-To-First-Token (28.8ms vs 32.2ms), it exacts a \textbf{38\% throughput penalty} to achieve it.

\begin{table}[h!]
\caption{The Backend Dichotomy: Qwen3-8B (NVFP4) Performance on RTX 5090 using TensorRT-LLM v1.1.0. All results are for the longest prompt variant (496 tokens)}
\label{tab:nvfp4-results}
\begin{center}
\begin{small}
\begin{sc}
\setlength{\tabcolsep}{8pt}
\begin{tabular}{llcc}
\toprule
\textbf{Framework} & \textbf{Backend} & \textbf{Throughput} & \textbf{TTFT} \\
& & (tok/s) & (ms) \\
\midrule
\multirow{2}{*}{TensorRT-LLM} & PyTorch (Default) & \textbf{151.4} & 32.2 \\
& C++ Engine (Legacy) & 93.1 & \textbf{28.8} \\
\midrule
llama.cpp & GGUF (BF16) & 92.2 & 44.0 \\
\bottomrule
\end{tabular}
\end{sc}
\end{small}
\end{center}
\end{table}

\textbf{Implication:} 
Developers are currently forced into a binary choice: utilize the PyTorch backend to access the actual speed benefits of NVFP4, or use the Legacy backend which negates nearly all throughput advantages of the new format.

\subsection{Architectural Efficiency: Dense vs. Mixture-of-Experts on Apple Silicon}

Table \ref{tab:large-models} presents the inference throughput across three distinct model architectures: a large-scale Mixture-of-Experts (MoE) (Qwen3-Next-80B), a dense model (Llama-3.3-70B), and a lightweight MoE (GLM-4.7-Flash-30B). While the RTX 5090 demonstrates superior raw throughput when models fit entirely within VRAM (e.g., $151.3$ tok/s for GLM-4.7), its performance degrades precipitously when offloading is required (dropping to $8.1$ tok/s for Qwen3), highlighting the ``PCIe Wall'' inherent to discrete consumer GPUs.

A more nuanced divergence appears when analyzing Apple Silicon performance, particularly the relationship between the M3 Ultra and the M4 Pro.

\paragraph{Bandwidth-Bound Regime (Dense Models):}
For the dense Llama-3.3-70B model, performance scales linearly with memory bandwidth. The M3 Ultra ($13.1$ tok/s) outperforms the M4 Pro ($5.1$ tok/s) by a factor of approximately $2.5\times$. This aligns with the hardware specifications, where the M3 Ultra's memory bandwidth ($\approx 800$ GB/s) is roughly triple that of the M4 Pro ($\approx 273$ GB/s). Since dense models must load all parameters for every token generated, the unified memory bandwidth acts as the strictly limiting factor.

\paragraph{Latency-Bound Regime (MoE Models):}
Conversely, the results for MoE models (Qwen3-Next and GLM-4.7) defy purely bandwidth-based expectations. In the case of Qwen3-Next-80B, the M4 Pro ($52.3$ tok/s) actually surpasses the M3 Ultra ($49.1$ tok/s). This inversion suggests a shift from a bandwidth-bound to a latency-sensitive regime.

MoE architectures activate only a subset of parameters per token (e.g., $\textit{active parameters} \ll \textit{total parameters}$), drastically reducing the memory bandwidth requirement. Once the bandwidth bottleneck is alleviated, inference speed becomes dependent on compute efficiency and memory access latency. We hypothesize that two factors contribute to the M4 Pro's advantage here:

\begin{enumerate}
    \item \textbf{Monolithic vs. Multi-Die Latency:} The M3 Ultra relies on the UltraFusion interconnect to bridge two M3 Max dies. The irregular memory access patterns inherent to MoE routing likely incur a latency penalty when traversing this interconnect, whereas the monolithic M4 Pro benefits from uniform, low-latency memory access.
    \item \textbf{Architectural Improvements:} The M4 architecture features improved single-core CPU performance and NPU enhancements. Since MoE inference involves complex routing logic (gating networks) that is often compute-bound at batch size 1, the newer architecture provides a throughput gain that raw bandwidth cannot compensate for.
\end{enumerate}

These findings indicate that for researchers running modern MoE models locally, newer microarchitectures (M4) may offer superior value to older, higher-bandwidth tiers (M3 Ultra), whereas dense model inference remains strictly the domain of high-bandwidth interconnects.

\begin{table*}[h]
\caption{High-Fidelity Model Inference: The Trade-off Between VRAM Constraints and Unified Memory. We compare distinct strategies: Nvidia's "Fit in VRAM" (aggressive quantization or smaller models), Nvidia's "Offload" (PCIe bottleneck), and Apple's "Native Capacity." Throughput is reported for the generation phase on long prompts. \textbf{OOM} indicates Out Of Memory.}
\label{tab:large-models}
\begin{center}
\begin{small}
\begin{sc}
\footnotesize 
\setlength{\tabcolsep}{12pt}
\begin{tabular}{lccc}
\toprule
\textbf{Device} & \textbf{Format} & \textbf{Offload} & \textbf{Throughput} \\ & & & (tok/s) \\
\midrule
\multicolumn{4}{c}{\textbf{GLM-4.7-Flash-30B (MoE)}} \\
\midrule
RTX 5090 & GGUF Q6\_K\_XL & 0\% & \textbf{151.3} \\
M4 Pro   & MLX 4-bit & 0\% & 53.6 \\
M3 Ultra & MLX 4-bit & 0\% & 55.0 \\
M2 Max   & MLX 4-bit & 0\% & 39.7 \\
\midrule
\multicolumn{4}{c}{\textbf{Llama-3.3-70B-Instruct (Dense)}} \\
\midrule
RTX 5090 & GGUF IQ3\_XXS & 0\% & \textbf{46.5} \\
RTX 5090 & GGUF Q4\_K\_M & 25\% (CPU) & 2.3 \\
M4 Pro   & MLX 4-bit & 0\% & 5.1 \\
M3 Ultra & MLX 4-bit & 0\% & 13.1 \\
M2 Max   & MLX 4-bit & N/A & OOM \\
\midrule
\multicolumn{4}{c}{\textbf{Qwen3-Next-80B (MoE)}} \\
\midrule
RTX 5090 & GGUF Q2\_K\_XL & 0\% & \textbf{76.1} \\
RTX 5090 & GGUF Q4\_K\_M & 28\% (CPU) & 4.7 \\
M4 Pro & MLX 4-bit & 0\% & 52.3 \\
M3 Ultra & MLX 4-bit & 0\% & 49.1 \\
M2 Max   & MLX 4-bit & N/A & OOM \\
\bottomrule
\end{tabular}
\end{sc}
\end{small}
\end{center}
\end{table*}

\section{Discussion}
\label{sec:discussion}

Our results paint a picture of two distinct, highly specialized ecosystems where the "optimal" choice is no longer defined by a single metric, but by the specific constraints of the workload: model size, required precision, and deployment latency.

\subsection{The VRAM Wall and the Intelligence-Throughput Trade-off}
Perhaps the most critical finding for practitioners is the behavior of the hardware at the memory limit. As demonstrated in Table~\ref{tab:large-models}, the Nvidia RTX 5090 offers unrivaled performance for models that fit entirely within its 32GB frame—achieving 151 tok/s on the 30B parameter GLM-4.7 compared to just 55 tok/s on the M3 Ultra.

However, for state-of-the-art 70B+ models, the "VRAM Wall" forces a stark trade-off between \textit{intelligence} and \textit{speed}. To run a 70B model on the RTX 5090, users must choose between:
\begin{enumerate}
    \item \textbf{Aggressive Quantization (IQ3/Q2):} This fits the model in VRAM, maintaining high throughput (76 tok/s), but relies on precisions known to degrade reasoning capabilities in complex tasks.
    \item \textbf{CPU Offloading (Q4):} Retaining the "smart" Q4 weights requires offloading $\approx25\%$ of the model. Our data shows this incurs a catastrophic penalty, dropping throughput to $\approx2-4$ tok/s due to the PCIe bus bottleneck.
\end{enumerate}

In contrast, the Apple ecosystem avoids this cliff entirely. The M3 Ultra runs the more precise 4-bit weights at a usable 13-49 tok/s (depending on architecture). This identifies the Unified Memory Architecture as convenient, usable, and scalable alternative to scaling up the amount of discrete GPUs in the node, which can be costly and complex to set up.

\subsection{Ecosystem Friction: The Hidden Cost of NVFP4}
While Nvidia's NVFP4 format delivers on its promise of "4-bit speed with FP16 accuracy", it introduces significant ecosystem friction. 

\textbf{The Backend Trap:} We identified a "Backend Dichotomy" in TensorRT-LLM v1.1.0. Users utilizing the legacy C++ engine sacrifice nearly all throughput gains of NVFP4 (93.1 tok/s), while the modern PyTorch backend achieves 151.4 tok/s but incurs a slightly higher startup latency. This requires users to benchmark internal framework components to realize the hardware's potential.

\textbf{The Creation Barrier:} Furthermore, we encountered a hard deployment wall: while an RTX 5090 can \textit{run} a quantized model, it often lacks the VRAM to \textit{create} one. Quantizing a 70B parameter model to NVFP4 requires peak memory far exceeding 32GB. This forces users to rely on slow CPU-based quantization (taking hours) or depend on pre-quantized weights (if exist).

\subsection{MoE Dynamics and Die Monolithicity}
Our finding that the M4 Pro outperforms the M3 Ultra on MoE workloads (Figure~\ref{tab:large-models}) chips away at the assumption that "more cores is always better." MoE models, which are latency-sensitive due to their routing logic, appear to suffer from the die-to-die interconnect overhead of the M3 Ultra's dual-chip design. This suggests that for the growing class of efficient MoE models (e.g., GLM, Qwen-MoE), modern monolithic architectures (M4) offer superior value to older, multi-die systems.

\section{Conclusion}
\label{sec:conclusion}

In this work, we conducted a systematic empirical comparison of consumer-grade Nvidia and Apple Silicon hardware for local LLM inference, extending our analysis from small 1.5B models to state-of-the-art 80B parameter workloads.

Our findings reveal a fundamental dichotomy. Nvidia's discrete GPUs, empowered by the new NVFP4 precision, establish the \textbf{Compute Density Frontier}. For models that fit within the VRAM buffer (typically $<30$B parameters), the RTX 5090 is the unequivocal choice, delivering up to $3\times$ the throughput of Apple Silicon. However, this performance is gated by significant "ecosystem friction," requiring users to navigate complex backend choices and VRAM constraints during model compilation.

Conversely, Apple's M-series SoCs define the \textbf{Memory Capacity Frontier}. By eliminating the PCIe bottleneck, the Unified Memory Architecture allows users to scale to 70B and 80B parameter models at practical 4-bit precisions. While lacking the raw burst speed of Nvidia's flagship, Apple's ecosystem offers a consistent, "offload-free" experience that prioritizes model fidelity over raw token generation speed.

Crucially, we identify a third dimension: the \textbf{Performance-Efficiency Continuum}. Our measurements uncover a stark disparity in operational cost, with the Apple M3 Ultra achieving up to $23\times$ higher energy efficiency (tokens/joule) compared to the RTX 5090. While Nvidia prioritizes peak throughput at the expense of power draw, Apple's SoC design offers a sustainable path for "always-on" local intelligence, where thermal envelopes and battery life are limiting factors.

Ultimately, the "Silicon Showdown" is no longer about which chip is faster. It is a strategic choice between the specialized, high-velocity compute of Nvidia for agentic and low-latency workflows, and the massive, unified capacity of Apple Silicon for running the world's most intelligent open-weights models locally.

%
%
%
\bibliographystyle{splncs04}
\bibliography{bibliography}




\end{document}